\documentclass[twocolumn,english,aps,prb,reprint,superscriptaddress,citeautoscript]{revtex4-1}
\usepackage[T1]{fontenc}
\usepackage{xcolor}
\usepackage{amsmath}
\usepackage{graphicx}
\PassOptionsToPackage{normalem}{ulem}
\usepackage{ulem}
\usepackage{subscript}
\usepackage{amsmath}
\usepackage{siunitx}
\usepackage{textgreek}
\usepackage{mhchem}
\usepackage{babel}

\makeatletter

\renewcommand*{\fnum@figure}{{\normalfont\bfseries \figurename~\thefigure}}
\renewcommand*{\@caption@fignum@sep}{\textbf{. }}
\addto\captionsenglish{\renewcommand{\figurename}{Fig.}}

\makeatother

\begin{document}

\title{Coherent Helicity-Dependent Spin-Phonon Oscillations in the Ferromagnetic van der Waals Crystal \ce{CrI3}}

\author{P. \surname{Padmanabhan}}
\thanks{Authors to whom correspondence should be addressed: prashpad@lanl.gov, arun.paramekanti@physics.utoronto.ca, jxzhu@lanl.gov, rpprasan@lanl.gov}
\affiliation{Center for Integrated Nanotechnologies, Los Alamos National Laboratory, Los Alamos, NM 87545, USA}

\author{F. L. \surname{Buessen}}
\affiliation{Department of Physics, University of Toronto, Toronto, Ontario M5S1A7, Canada}

\author{R. \surname{Tutchton}}
\affiliation{Center for Integrated Nanotechnologies, Los Alamos National Laboratory, Los Alamos, NM 87545, USA}

\author{K. W. C. \surname{Kwock}}
\affiliation{Center for Integrated Nanotechnologies, Los Alamos National Laboratory, Los Alamos, NM 87545, USA}
\affiliation{The Fu Foundation School of Engineering and Applied Science, Columbia University, New York, NY 10027, USA}

\author{S. \surname{Gilinsky}}
\affiliation{Center for Integrated Nanotechnologies, Los Alamos National Laboratory, Los Alamos, NM 87545, USA}

\author{M.-C. \surname{Lee}}
\affiliation{Center for Integrated Nanotechnologies, Los Alamos National Laboratory, Los Alamos, NM 87545, USA}

\author{M. A. \surname{McGuire}}
\affiliation{Materials Science and Technology Division, Oak Ridge National Laboratory, Oak Ridge, TN 37831, USA}

\author{S. R. \surname{Singamaneni}}
\affiliation{Department of Physics, The University of Texas at El Paso, El Paso, TX 79968, USA}

\author{D. A. \surname{Yarotski}}
\affiliation{Center for Integrated Nanotechnologies, Los Alamos National Laboratory, Los Alamos, NM 87545, USA}

\author{A. \surname{Paramekanti}}
\thanks{Authors to whom correspondence should be addressed: prashpad@lanl.gov, arun.paramekanti@physics.utoronto.ca, jxzhu@lanl.gov, rpprasan@lanl.gov}
\affiliation{Department of Physics, University of Toronto, Toronto, Ontario M5S1A7, Canada}

\author{J.-X. \surname{Zhu}}
\thanks{Authors to whom correspondence should be addressed: prashpad@lanl.gov, arun.paramekanti@physics.utoronto.ca, jxzhu@lanl.gov, rpprasan@lanl.gov}
\affiliation{Center for Integrated Nanotechnologies, Los Alamos National Laboratory, Los Alamos, NM 87545, USA}

\author{R. P. \surname{Prasankumar}}
\thanks{Authors to whom correspondence should be addressed: prashpad@lanl.gov, arun.paramekanti@physics.utoronto.ca, jxzhu@lanl.gov, rpprasan@lanl.gov}
\affiliation{Center for Integrated Nanotechnologies, Los Alamos National Laboratory, Los Alamos, NM 87545, USA}

\maketitle

\textbf{The discovery of two-dimensional (2D) systems hosting intrinsic long-range magnetic order \cite{Burch2018} represents a seminal addition to the rich physical landscape of van der Waals (vdW) materials. \ce{CrI3} has emerged as perhaps the most salient example, as the interdependence of crystalline structure and magnetic order \cite{Sivadas2018}, along with strong light-matter interactions \cite{Seyler2018,Sun2019} provides a promising platform to explore the optical control of magnetic, vibrational, and charge degrees of freedom at the 2D limit. However, the fundamental question of how this relationship between structure and magnetism manifests on their intrinsic timescales has rarely been explored. Here, we use ultrafast optical spectroscopy to probe magnetic and vibrational dynamics in \ce{CrI3}, revealing demagnetization dynamics governed by spin-flip scattering and remarkably, a strong transient exchange-mediated interaction between lattice vibrations and spin oscillations. The latter yields a coherent spin-coupled phonon mode that is highly sensitive to the helicity of the driving pulse in the magnetically ordered phase. Our results shed light on the nature of spin-lattice coupling in vdW magnets on ultrafast timescales and highlight their potential for applications requiring non-thermal, high-speed control of magnetism at the nanoscale.}

Since its first demonstration \cite{Beaurepaire1996}, the ultrafast manipulation of magnetism has been a major topic of research, with significant efforts aimed at unraveling the nature of dynamic demagnetization \cite{Koopmans2010,Afanasiev2019}, the excitation of collective magnetic modes \cite{Padmanabhan2019,Tzschaschel2019}, and the realization of all-optical switching \cite{Mangin2014,Stupakiewicz2017,Schlauderer2019}. The recent emergence of 2D materials hosting long-range magnetic order \cite{Gong2017,Huang2017} presents a new, relatively unexplored playground to investigate these phenomena in systems where the interplay between structural order and exchange interactions plays a pivotal role. \ce{CrI3} is a prototypical example of such a system, hosting out-of-plane ferromagnetic (FM) order down to the monolayer limit \cite{Huang2017} stabilized by uniaxial anisotropy, which opens a gap in the low energy zone-center magnon spectrum \cite{Lee2020}. However, for even numbers of atomic layers the relationship between stacking order and Cr-Cr superexchange interactions \cite{Sivadas2018} stabilizes an antiferromagnetic (AFM) phase with broken inversion symmetry. This gives rise to unique nonlinear optical phenomena \cite{Sun2019} and highlights the strong influence of the crystal structure (Fig. 1(a)) on the magnetic order and optical response of \ce{CrI3}. Additionally, recent inelastic neutron scattering \cite{Chen2018} and Raman spectroscopic \cite{Jin2018,Li2020} measurements reveal terahertz (THz) optical spin-wave branches that are close in energy to several vibrational modes in \ce{CrI3} \cite{Webster2018}, pointing to the possibility of dynamic coupling between the lattice and spin degrees of freedom \cite{Streib2019}.

\begin{figure}
\includegraphics[width=\linewidth]{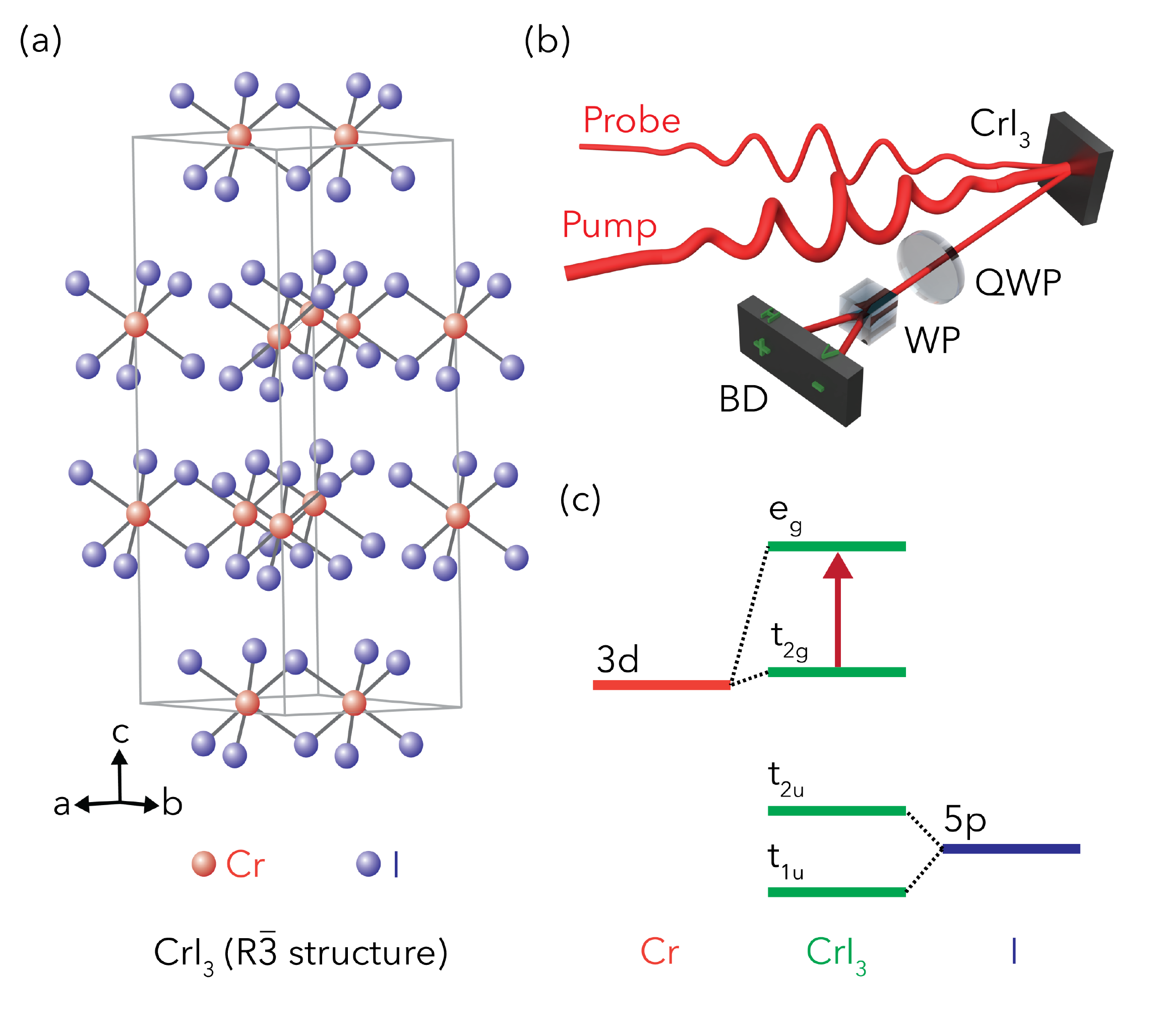}
\caption{\textbf{\ce{CrI3} structural and electronic properties and TRPR experimental scheme.} Schematics of (a) the crystal structure of the low temperature R$\bar{3}$ phase of \ce{CrI3}, (b) the femtosecond TRPR experiment, where QWP is a quarter wave plate, WP is a Wollaston prism, and BD is a balanced detector, and (c) the relevant energy levels in \ce{CrI3}, where the red arrow denotes optical transitions at $\sim 1.5\textrm{ \si{\eV}}$ that generate excited electrons and holes. Dashed lines indicate the dominant atomic orbital character of the ligand-field split levels. Transitions between the $e_g$, $t_{2u}$, and $t_{1u}$ levels significantly contribute to coherent phonon generation.}
\end{figure}

In this letter, we uncover the nature of this coupling by measuring the transient magnetization and coherent vibrational dynamics in \ce{CrI3} after femtosecond optical photoexcitation. We find that the demagnetization dynamics are driven by spin-flip scattering, while coherent pump helicity-dependent oscillations in the time-resolved polarization rotation (TRPR) signal highlight the strong influence of magnetic order on the $c$-axis $A_{1g}$ optical phonon mode. This new insight into dynamic spin-lattice coupling in vdW magnets opens a unique pathway for manipulating magnetic order through the vibrational degree of freedom, with significant implications for future nanoscale optomagnetic applications.

\begin{figure}
\includegraphics[width=\linewidth]{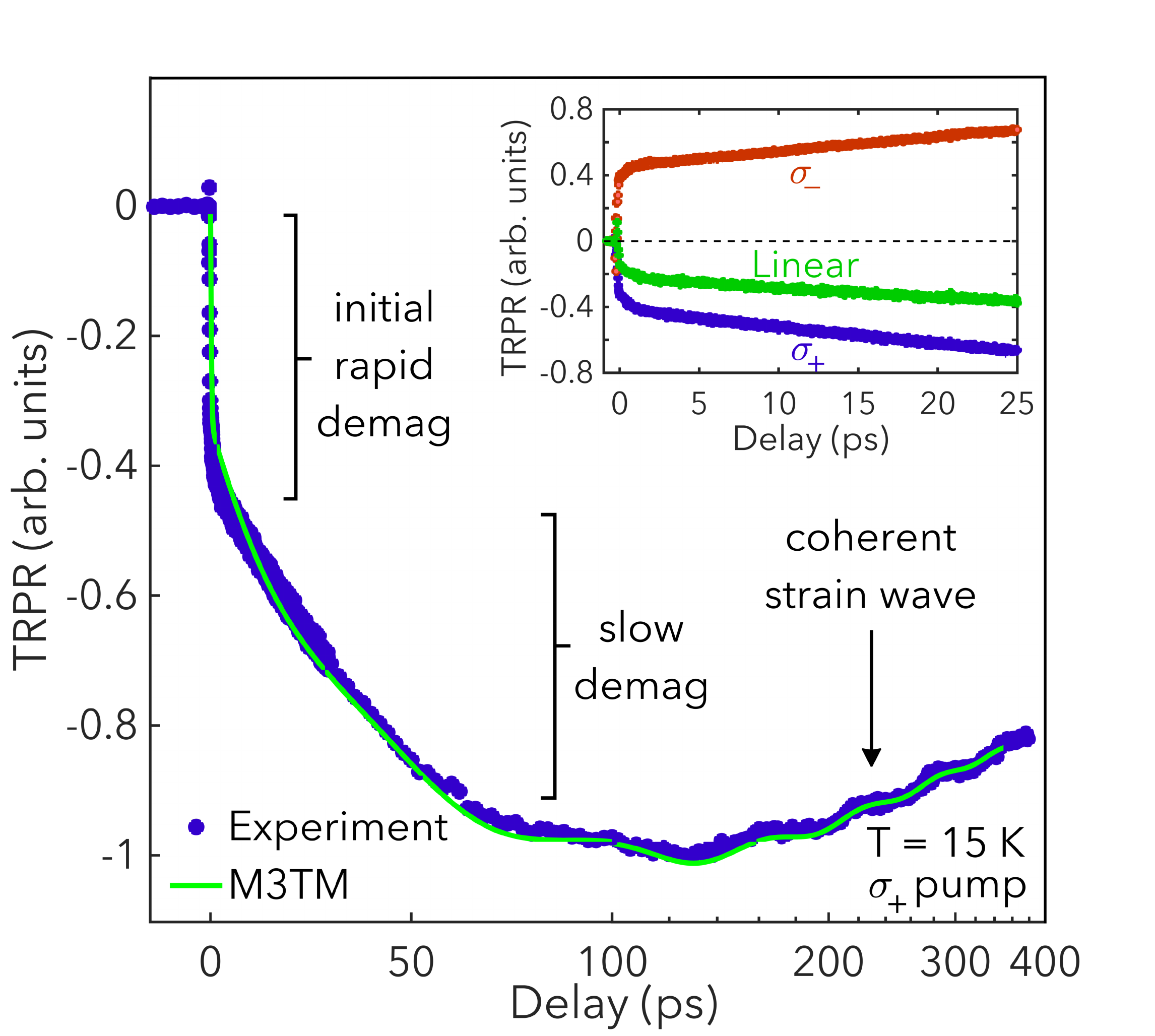}
\caption{\textbf{Ultrafast photoinduced demagnetization in \ce{CrI3}.} Time-resolved polarization rotation signal under $\sigma_+$ pumping at $T=\SI{15}{\si{\kelvin}}$, where the blue dots are experimental data and the green line is a fit using the M3TM (see SI). The oscillatory component in the M3TM fit curve was obtained by separately fitting the oscillatory component of the experimental data with a decaying sinusoidal function. The inset shows the magnetization dynamics on shorter timescales under $\sigma_+$, $\sigma_-$, and linearly polarized pumping at $T=\SI{15}{\si{\kelvin}}$.}
\end{figure}

In our experiments, 1.55 eV, 85 fs pump and probe pulses (measured at the sample position) were focused at near-normal incidence onto as-grown bulk-like flakes of \ce{CrI3} (Curie temperature, $T_C=\SI{61}{\si{K}}$)  (Fig. 1(b)). Fig. 2(a) shows the resulting TRPR signal obtained in the FM phase at $T=\SI{15}{\si{\kelvin}}$ under right circularly polarized ($\sigma_+$) pumping. We observe a rapid sub-picosecond decrease in the magnetization, a further reduction over tens of picoseconds, a $\sim\SI{16}{\si{GHz}}$ oscillation due to a coherent strain wave, and an eventual nanosecond timescale recovery. This two-step demagnetization process, referred to as type-II dynamics, has been observed in other ferromagnets \cite{Kim2009,Gunther2014} and occurs when demagnetization is not completed before electron-phonon equilibration \cite{Koopmans2010}. 

More insight can be obtained by considering the optical transitions in \ce{CrI3} following $\SI{1.55}{\electronvolt}$ excitation. The absorption spectrum of bulk \ce{CrI3} shows a weak resonance peak at $\sim\SI{1.5}{\electronvolt}$, attributed to transitions between the partially filled $t_{2g}$ and unfilled $e_g$ levels resulting from the octahedral ligand field-induced splitting of the \ce{Cr^3+} $d$-orbital (Fig. 1(c)). Despite the even parity of these states, this transition is allowed due to mixing with various odd-parity states \cite{Seyler2018}, and may also be associated with a low-lying bright charge-transfer exciton \cite{Wu2019b}. Regardless, below $T_C$ the $\SI{1.55}{\electronvolt}$ pump pulse drives vertical transitions predominantly involving majority-spin states, due to significant spin polarization of the valence and conduction bands \cite{,Wu2019b}. This leads to a strong preferential absorption of $\sigma_+$ light in spin-up domains at $\SI{1.55}{\electronvolt}$ \cite{Wu2019b}. 

The strong helicity-dependent absorptivity of \ce{CrI3} suggests that the ultrafast magnetic response should be nearly negligible for left circularly polarized ($\sigma_-$) pump pulses. However, the TRPR signal under $\sigma_-$ pumping has a sign opposite to the $\sigma_+$ case (Fig. 2 inset). This helicity-dependent response is preserved for all temperatures $T<T_C$. We attribute this to the presence of multiple domains within the photoexcited region \cite{McGuire2015,Zhong2017}; i.e., spin-down domains preferentially absorb $\sigma_-$ photons, leading to their subsequent demagnetization.  Nevertheless, horizontally polarized pumping yields a negative signal (Fig. 2 inset), implying that the volume fraction of spin-up domains is higher in the probed region. 

\begin{figure*}[t]
\includegraphics[width=\linewidth]{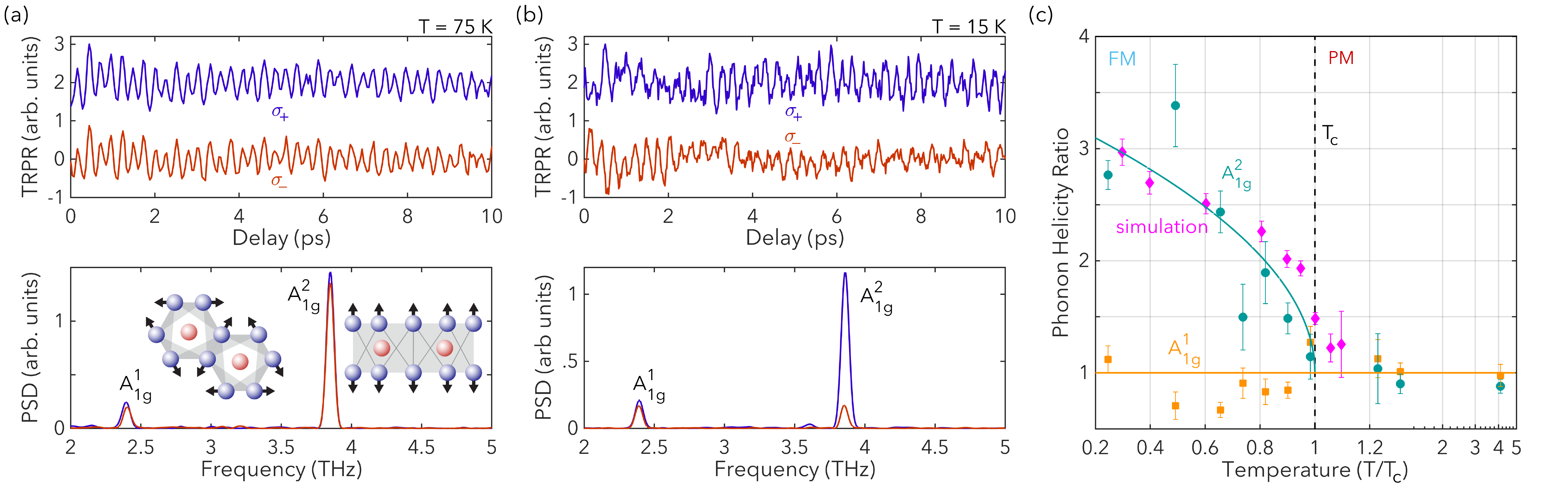}
\caption{\textbf{Coherent spin-coupled phonon dynamics as a function of temperature.} Oscillations in the time-domain signals (upper panel) after subtraction of the demagnetization background, under $\sigma_+$ (blue) and $\sigma_-$ (red) pumping, and their corresponding power spectral densities (PSD, lower panel) at (a) $T=\SI{75}{\si{K}}$ and (b) $T=\SI{15}{\si{K}}$. The inset of (a) shows a schematic of the eigenvectors associated with the two phonon modes. (c) The $\sigma_+$/$\sigma_-$ ratio of the integrated Fourier transform peaks of the measured signal at the $A^1_{1g}$ (orange) and $A^2_{1g}$ (green) mode frequencies, and the simulated helicity-dependent ratio at the $A^2_{1g}$ frequency (pink) vs. normalized temperature. The solid green line is a fit using a FM order-parameter-like function $\propto\sqrt{T_C-T}$ and the solid orange line is a guide to the eye. Error bars were obtained from a bootstrap sampling analysis.}
\label{fig:coherent_data_and_ratios}
\end{figure*}

Accordingly, upon absorption of a $\sigma_+$ pulse, majority-spin electrons and holes are excited in \ce{CrI3}. Due to spin-orbit coupling (SOC), the wave functions of these carriers are a mixture of pure spin states \cite{Carva2011}. This allows for a finite probability of Elliott-Yafet spin-flip scattering processes\cite{Steiauf2009}, which mediate spin relaxation, particularly in the hole population due to the comparatively small valence band exchange splitting \cite{Wu2019b}. Furthermore, the photoexcited carrier density of $\sim\SI{e19}{\si{\per\cubic\centi\meter}}$ exceeds the Mott density criterion, given the nanometer-scale exciton radius \cite{Wu2019b}, leading to a transient quasi-metallic state. This makes it possible to describe ultrafast dynamics in \ce{CrI3} using the microscopic three-temperature model (M3TM), in which the excited electronic system supplies the energy for demagnetization, while interactions with the lattice allow for angular momentum dissipation \cite{Koopmans2010}. A fit to the demagnetization dynamics with the M3TM (Fig. 2) accurately reproduces both demagnetization steps with a spin-flip probability of $a_{\textrm{sf}}=0.175$, consistent with other materials showing type-II dynamics \cite{Kim2009,Gunther2014}. More detail is included in the supplementary information (SI).

We now turn our attention to the coherent dynamics. The top panel of Fig. 3(a) shows the TRPR signal at $T=\SI{75}{\si{\kelvin}}>T_c$ under $\sigma_+$ and $\sigma_-$ pumping after subtracting the demagnetization background discussed above, revealing pronounced coherent oscillations. The power spectral density (PSD) in the bottom panel reveals two distinct modes at $\sim\SI{3.87}{\si{\THz}}$ and $\sim\SI{2.37}{\si{\THz}}$, corresponding to $c$-axis ($A^2_{1g}$) and in-plane ($A^1_{1g}$) Raman-active phonons, respectively \cite{Webster2018,Jin2018,Li2020}. These modes are excited via impulsive stimulated Raman scattering (ISRS), which is permitted by their $A_{1g}$ symmetry \cite{Li2020}.

Notably, the PSD of both modes is nearly identical for both pump helicities above $T_C$. However, this symmetry is broken below $T_C$, where $\sigma_-$ pumping leads to a smaller $A^2_{1g}$ amplitude (Fig. 3(b)). In contrast, the $A^1_{1g}$ mode remains relatively invariant to the choice of pump polarization. This can be seen in Fig. 3(c), where we plot the $\sigma_+$/$\sigma_-$ ratio of the integrated spectral peaks of both oscillatory modes vs. $T$. Here, the $A^2_{1g}$ ratio below $T_C$ consistently follows a FM order-parameter-like function, $\propto\sqrt{T_C-T}$, similar to the case of THz magnons \cite{Jin2018}. This highlights that the coherent amplitude of the $A^2_{1g}$ mode is especially sensitive to the underlying magnetic order in the system. In contrast, the $A^1_{1g}$ mode ratio shows minimal variation with temperature. This difference is striking, and suggests that the helicity dependence of the $A^2_{1g}$ mode does not originate from a dominant optical effect (i.e., the larger absorptivity of $\sigma_+$ light in \ce{CrI3} below $T_C$), as this would impact the two modes in a similar manner. Additionally, the diagonal Raman tensors associated with both $A_{1g}$ modes eliminates the possibility that the helicity-dependence originates from the form of the ISRS selection rule \cite{Jin2018}. 

Instead, to understand the clear sensitivity of the $A^2_{1g}$ mode to the underlying magnetic order, we recall that this mode corresponds to a $c$-axis oscillation of iodine atoms (lower panel of Fig. 3(a)), leading to an oscillatory trigonal distortion of the \ce{CrI6} octahedra. This distortion can transiently modulate spin exchange interactions through anisotropy- or exchange-mediated pathways \cite{Streib2019,Liu2020}. This is even more likely in \ce{CrI3}, given the close energetic proximity between the $A_{1g}$ phonon modes and Raman-active magnons \cite{Jin2018,Li2020}. We confirmed this using density functional theory (DFT) to calculate $\Delta E =  E_\mathrm{FM} - E_\mathrm{AFM}$, the energy difference per spin between FM and AFM configurations, which is a measure of the overall exchange strength. Tracking its dependence on the lattice displacement, $X$, of the $A^2_{1g}$ phonon yields $\partial\Delta E/ \partial X \approx -32\>\si{\meV/\angstrom}$ per \ce{Cr} atom, signaling the importance of spin-lattice coupling in \ce{CrI3}. Here, $X > 0$ denotes trigonal compression of the \ce{CrI6} octahedra.

To explain the helicity-dependent TRPR, we then developed a spin-phonon model for \ce{CrI3}. This describes $S\!=\! 3/2$ \ce{Cr} moments on an $N$-site honeycomb lattice interacting via nearest-neighbor exchange interactions and coupled to the global $A^2_{1g}$ phonon displacement. The total Hamiltonian is $H=H_{\mathrm{ph}}+H_{\mathrm{sp}}$, with the phonon term $H_\mathrm{ph} = P^2/2M+M\Omega^2X^2/2$. Here $P$ is the momentum, $M/N\approx6.3\times 10^{-25}~\mathrm{kg}$ is the mass of the three iodine atoms per \ce{Cr} spin, and $\Omega\approx\SI{3.87}{\si{\THz}}$ is the $A^2_{1g}$ phonon frequency. The spin Hamiltonian is
\begin{equation}
\! H_\mathrm{sp}\!=\!\!\!\!\! \sum\limits_{\langle \mathbf{r} \mathbf{r}' \rangle_\gamma} \!\!\! \left[ \widetilde{J}_H \mathbf{S}_\mathbf{r} \!\cdot\! \mathbf{S}_{\mathbf{r}'} 
\!+\! \widetilde{J}_K S_\mathbf{r}^\gamma S_{\mathbf{r}'}^\gamma 
\!+\! \widetilde{J}_\Gamma \big( S_{\mathbf{r}}^\alpha S_{\mathbf{r}'}^\beta \!+\! S_{\mathbf{r}}^\beta S_{\mathbf{r}'}^\alpha \big) \right]
\label{Eq: Hsp}
\end{equation}
where $\langle\mathbf{r}\mathbf{r}'\rangle$ denotes nearest-neighbor sites, $\widetilde{J}_H$ is a Heisenberg coupling, and the bond-anisotropic interactions $\widetilde{J}_K$ and $\widetilde{J}_\Gamma$ correspond to Kitaev and off-diagonal symmetric exchange, respectively \cite{Lee2020}. The superscript $\gamma$ denotes the direction of the bond, with $(\alpha,\beta) \perp \gamma$.  The weaker interlayer couplings in bulk \ce{CrI3} \cite{Chen2018} are not expected to qualitatively impact our results, justifying this monolayer model. The $A^2_{1g}$ lattice distortion modifies the exchange constants as $\widetilde{J}_i \!=\! J_i (1\!+\! \phi_i X)$, with spin-phonon coupling parameters $\phi_i$ ($i=H,K,\Gamma$). We fix $(J_H,J_K,J_\Gamma)\!=\! (-1.3,-3.0,-0.5)~\mathrm{meV}$ to reasonably capture the spin-wave dispersion \cite{Chen2018,Lee2020} and the monolayer $T_C$, and set $(\phi_H,\phi_K) \!=\! (6, -3)~\si{\angstrom}^{-1}$ to reproduce our DFT results for $(\partial \Delta E/\partial X)$ (see SI). For $T \ll T_C$, our model predicts an equilibrium displacement $\langle X \rangle \!\sim\!5\times10^{-4}\si{\angstrom}$, comparable to the experimentally estimated magnetostriction \cite{Jiang2020}.

Using this spin-phonon model, we simulate the impact of the pump pulse as an instantaneous helicity-dependent lattice distortion, $X_\pm(t\!=\! 0) \!=\! X+\xi_1\left( 1 \pm \xi_2 m \right)$. Here, $X$ and $m$ are the equilibrium lattice displacement and magnetization respectively, and the signs $\pm$ correspond to the $\sigma_{\pm}$ pump helicity. $\xi_1$ determines the overall strength of the distortion due to the Raman process, and $\xi_2$ represents its helicity dependence; the latter can arise from the spin selectivity of the helicity-dependent photoexcitation, which transiently enhances or suppresses the local Cr moment \cite{Ron2019}. We solve for the subsequent dynamics of the spin-phonon system by numerically integrating the coupled equations of motion (see Methods). For $T>T_C$, we find persistent oscillations in $X_\pm(t)$ at the phonon frequency $\Omega$, while the uniform magnetization $m_{{\pm}} (t)$ does not exhibit coherent dynamics. Remarkably, in the FM phase ($T<T_C$), the distortion leads to coupled, coherent oscillations in $X_\pm (t)$ and the magnetization $m_{_\pm}(t)$ at $\Omega$ (see SI). The coherent $m_{\pm}(t)$ oscillations lead to oscillations in the polarization rotation, creating an additional temperature and helicity-dependent contribution to the TRPR signal below $T_C$. Setting $\xi_2 = 0.18/\si{\mu_B}$, we find that $m_{_+}(\Omega)/m_{_-}(\Omega)$ plotted vs. $T/T_C$ shows excellent agreement with the experimental TRPR ratio below $T_C$ (Fig. 3(c)).

Ultimately, the overarching physical mechanism underlying the helicity-dependent response of the $A^2_{1g}$ mode centers upon the coherent coupling between the magnetization and the $A^2_{1g}$ phonon. Above $T_C$, the femtosecond optical pump drives transitions involving the Cr-like $e_g$ level. The partial occupation of this level leads to a strongly Jahn-Teller active ion, causing the system to undergo an ultrafast trigonal distortion that triggers the coherent phonon mode. Below $T_C$, the modulation of the spin exchange by this phonon mode leads to the intertwining of lattice vibrations and coherent spin oscillations. Moreover, transitions from strongly iodine-like states (e.g., $T_{1u}$ and $T_{2u}$, see Fig.1(c)) lead to helicity-dependent changes in the local magnetization, transiently enhancing (suppressing) it under $\sigma_+$ ($\sigma_-$) photoexcitation in a majority spin-up photoexcited volume. As shown above, this contributes a helicity-dependent component to the impulsive lattice distortion, akin to a dynamic magnetostriction process. 

In conclusion, our measurements and simulations highlight the inextricable link between magnetic order and structure in \ce{CrI3}, due to the intimate relationship between the strength of exchange interactions, trigonal lattice distortions, and the orbital character of the states involved in the photoexcitation process. This in turn allows for greater flexibility in the manipulation of magnetization and vibrational dynamics in 2D materials, especially using all-optical techniques that exploit nonlinear processes to drive coherent phenomena. Finally, our results shed new light on the immense potential for \ce{CrI3} and similar 2D magnetic materials in the next generation of optomagnetic technologies, and also provide insight into enabling ultrafast optical control of magnetism at the atomic limit.

\section{Methods}
\subsection{Time-resolved experiments}
Bulk-like flakes of \ce{CrI3}, grown using chemical vapor transport, were deposited onto an Au coated sample holder and placed in a variable temperature liquid helium flow optical cryostat. Ultrafast pulses were supplied by a regeneratively amplified Ti:Sapphire laser, which generated pulses with an 800 nm central wavelength, 85 fs duration (measured at the sample position), and 100 kHz repetition rate. The beam was split into pump and probe arms using a plate beam splitter. The pump beam was sent through a mechanical delay line, and both beams were passed through achromatic wave plates and subsequently focused onto optically flat regions of the sample at near-normal incidence using a 20X near-IR optimized apochromatic objective. The nominal $1/e^2$ focal spot diameter of the probe was approximately $\SI{10}-\SI{15}{\si{\um}}$, and the pump was approximately $\SI{20}{\si{\um}}$, yielding a pump fluence of $\sim\SI{0.64}{\si{\milli\joule\per\square\cm}}$. We performed experiments where the pump polarization was horizontally polarized (i.e., parallel to the optical table surface), right circularly polarized, and left circularly polarized. The probe beam was linearly polarized, and upon reflection, the beam was decomposed into its two orthogonal linear components using a Wollaston prism. These components were then differentially detected using a balanced Si photodiode detector and the output signal was fed into a lock-in amplifier to isolate the TRPR signal. Both the pump and probe beams were modulated using a 7/5 slotted optical chopper, allowing the lock-in amplifier to be referenced to the sum inter-modulation frequency (1.2 kHz) to eliminate pump scattering effects.

\subsection{Density functional theory calculations}
Our {\it ab initio} simulations were carried out within density functional theory. We used both the pseudopotential plane wave method implemented in the Vienna ab initio simulation package (VASP)~\cite{Kresse1993,Kresse1996} and in Quantum Espresso (QE), and the full-potential all-electron linearized augmented plane wave (FP-LAPW) method implemented in the Wien2k code~\cite{Blaha2020}.  The local density approximation was used for the exchange-correlation functional throughout all our {\it ab initio} simulations.  For the VASP calculations, we chose an energy cutoff of 500 eV for the plane-wave basis set and used a $10\times 10 \times 10$ ($10 \times 10 \times 3$) $\Gamma$-centered $k$-point mesh to sample the bulk (monolayer) Brillouin zone to perform the structure optimization, $\Gamma$-point phonon, Raman tensor (aided with the vasp\_raman python script~\cite{Fonarim2013}), and electronic structure calculations. The QE calculations used a plane-wave basis set with a cutoff energy of 60 Ry. A $10\times 10 \times 10$ $k$-point mesh was used to calculate the electronic structure of bulk CrI$_{3}$ and to support a calculation of the phonon dispersion at the $\Gamma$-point. In the Wien2k simulations, we focused on calculating the electronic band structure for bulk CrI$_3$  by taking a $15 \times 15 \times 15$ $k$-point mesh with a muffin-tin radius of 2.5$a_0$ (Cr) and 2.35$a_0$ (I). Here $a_0$ is the Bohr radius. 

We also used VASP to perform total energy calculations for the FM and AFM states of a CrI$_3$ monolayer, from which a Heisenberg model with nearest-neighbor exchange interactions was fit to obtain the exchange interaction. The spin-lattice coupling strength was then determined by distorting the equilibrium lattice according to the displacement field for a chosen vibration mode. The electronic structure obtained in our simulations on both bulk and 2D CrI$_3$ is consistent with those reported in the literature \cite{Webster2018,KumarGudelli2019,Lado2017}. We note that SOC reduces the band gap. However, since SOC does not qualitatively change the lattice dynamics \cite{Webster2018}, we did not include it in the calculation of the Raman tensor and estimate of the spin-lattice coupling strength.

\subsection{Monte Carlo simulations}
We performed classical Monte Carlo simulations on the coupled spin-phonon system described by the Hamiltonian $H=H_{\rm ph}+H_{\rm sp}$ to generate a thermal ensemble of states. We studied lattices of $2 \times L\times L$ spins, with system sizes up to $L=48$, with periodic boundary conditions. To improve the statistical convergence of the simulations, we employed a parallel tempering scheme~\cite{Hukushima1996} to simulate 144 logarithmically spaced temperature points between $T_\mathrm{min}\approx 7~\mathrm{K}$ and $T_\mathrm{max}\approx 90~\mathrm{K}$ in parallel. Each simulation was equilibrated for $10^6$~sweeps before taking measurements of the specific heat and the equilibrium magnetization for an additional $10^7$~sweeps; a single sweep is defined as one attempted update per spin or phonon degree of freedom. As shown in the SI, the chosen coupling constants reasonably reproduce the experimental spin-wave dispersion.~\cite{Chen2018} In addition, the computed temperature-dependent specific heat and magnetization yield a $T_C$ in reasonable agreement with experiments.

\subsection{Dynamical simulations}
We modeled the impact of the pump laser as a helicity-dependent lattice distortion, given by $X_{\pm}(t=0)= X + \xi_1 \left( 1 + \sigma \xi_2 m \right)$ at time $t=0$. Here, $X$ is the equilibrium lattice displacement, $m$ is the equilibrium magnetization of the spin configuration, and the sign $\pm$ distinguishes $\sigma_+$ and $\sigma_-$ pump pulses. Such a distortion could arise from an ultrafast trigonal splitting of a photoexcited Jahn-Teller active Cr $e_g$ level, with the splitting being sensitive to the local Cr spin due to Hund's coupling. We applied this ultrafast distortion to individual configurations selected from our Monte Carlo ensemble with $L=32$. The post-distortion dynamics were described using the Landau-Lifshitz equations for spins,
\begin{align}
\label{eq:landaulifshitz}
\hbar \frac{\mathrm{d}S_\mathbf{r}^x}{\mathrm{d}t} \!=\!\!\!\! &\sum\limits_{\gamma=x,y,z} \!\!\!\widetilde{J}_H \! \left( S_\mathbf{r}^z S_{\mathbf{r}_x}^z 
\!-\! S_\mathbf{r}^y S_{\mathbf{r}_x}^z \right) 
\!+\! \widetilde{J}_K \! \left( S_\mathbf{r}^z S_{\mathbf{r}_y}^y \!-\! S_\mathbf{r}^y S_{\mathbf{r}_z}^z \right) \nonumber\\
& \!+\! \widetilde{J}_\Gamma \left( S_\mathbf{r}^z S_{\mathbf{r}_x}^z \!+\! S_\mathbf{r}^z S_{\mathbf{r}_z}^z \!-\! S_\mathbf{r}^y S_{\mathbf{r}_x}^y \!-\! S_\mathbf{r}^y S_{\mathbf{r}_y}^x \right) \,,
\end{align}
with cyclic permutations of $(x,y,z)$ yielding $\mathrm{d}S_\mathbf{r}^{y}/\mathrm{d}t$ and $\mathrm{d}S_\mathbf{r}^{z}/\mathrm{d}t$, coupled with Newton's equations for the lattice
\begin{equation}
\frac{\mathrm{d}X}{\mathrm{d}t} = \frac{P}{M} \quad\textrm{and}\quad
\frac{\mathrm{d}P}{\mathrm{d}t} = -M \Omega^2 X - \frac{\partial H_\mathrm{spin}}{\partial X} \,.
\label{eq:Newton}
\end{equation}
In equation~\eqref{eq:landaulifshitz}, $\mathbf{r}_\gamma$ denotes the nearest-neighbor lattice site of $\mathbf{r}$ along bond direction $\gamma$. These equations were numerically integrated using a ninth-order Runge-Kutta algorithm~\cite{Rackauckas2017} and adaptive time steps with a local relative error tolerance of $\varepsilon_\mathrm{rel}=10^{-13}$. This yielded converged results up to timescales of approximately $25\,\mathrm{ps}$ in the magnetically ordered phase. We averaged the resulting dynamical observables over $2000$ Monte Carlo configurations with positive magnetization (when projected onto the twofold degenerate polarization axis). In particular, we extracted the time-dependent deviation of the magnetization, $\Delta m_{\pm}(t) = \langle m_{\pm}(t) \rangle - m$, where $m$ is the mean magnetization in thermal equilibrium and $\langle m_{\pm}(t) \rangle$ is the time-dependent magnetization after the ultrafast $\sigma_\pm$-helicity-dependent distortion was applied. We determined the ratio of Fourier components $m_{+}(\Omega)/m_{-}(\Omega)$ by calculating the Fourier transform $m_{\pm}(\Omega) = \mathrm{FT}(\Delta m_{\pm}(t))$ and fitting a Gaussian profile to the peak found at the $A_{1g}^{2}$ phonon frequency $\Omega$. The ratio $m_{+}(\Omega)/m_{-}(\Omega)$ is independent of the overall distortion $\xi_1$, while a helicity-dependent splitting $\xi_2 \approx 0.18 /\si{\mu_B}$ was found to reproduce our experimental results below $T_C$. Additional results on the time-resolved magnetization and phonon displacement are shown in the SI. Finally, our results are qualitatively robust upon tuning the Hamiltonian parameters, as long as there are at least two exchange interactions with distinct spin-phonon couplings.

\begin{acknowledgments}
P.P. and R.P.P. thank Stuart A. Trugman and Roberto Merlin for many insightful discussions. We gratefully acknowledge the support of the U.S. Department of Energy through the Los Alamos National Laboratory LDRD Program under project \#20190026DR. S.R.S acknowledges support from a UTEP start-up grant. AP and FLB acknowledge funding from NSERC of Canada. Crystal growth and characterization at Oak Ridge National Laboratory was supported by the US Department of Energy, Office of Science, Basic Energy Sciences, Materials Sciences and Engineering Division. M.C.L. also acknowledges support from the US Department of Energy, Office of Science, Basic Energy Sciences, Materials Sciences and Engineering Division. The Monte Carlo and dynamical simulations were performed on the Cedar cluster, hosted by WestGrid and Compute Canada. This work was performed, in part, at the Center for Integrated Nanotechnologies, an Office of Science User Facility operated for the U.S. Department of Energy (DOE) Office of Science, under user proposals \#2018BU0010 and \#2018BU0083. Los Alamos National Laboratory, an affirmative action equal opportunity employer, is managed by Triad National Security, LLC for the U.S. Department of Energy’s NNSA, under contract 89233218CNA000001.
\end{acknowledgments}

\section*{Author Contributions}
S. R. S., P.P, and R.P.P. conceived, initiated, and designed the experiment. P.P. built the experimental setup and carried out the measurements, assisted by K.W.C.K and S.G. P.P. and K.W.C.K. analyzed and interpreted the data with input from R.P.P., M.C.L. and D.A.Y. and theoretical support from J.-X.Z., R.T., A.P., and F.L.B. Crystals were grown by M.A.M. and initial characterization was performed by S.R.S. and M.A.M. J.-X.Z. and R.T. carried out DFT calculations. F.L.B. and A.P. conducted the Monte Carlo and dynamical simulations. P.P., R.P.P, F.L.B, and A.P. wrote the manuscript with input from all co-authors. 

\section*{Competing Interests}
The authors declare no competing interests.

\end{document}